\documentclass[nofootinbib,prd,aps,onecolumn,preprintnumbers,amsmath,amssymb,superscriptaddress]{revtex4}
\usepackage{graphicx} 
\usepackage{amssymb}
\usepackage{amsmath}
\usepackage[colorlinks=true]{hyperref}
\usepackage{ulem}

\begin{document}

\title{On Avoidance of PBH overproduction in inflation model with modified dispersion relation}

\author{Chengrui Yang}
\email{ycrd@hust.edu.cn}
\affiliation{School of Physics, Huazhong University of Science and Technology, Wuhan, 430074, China}

\author{Weixin Cai}
\email{weixincai@hust.edu.cn}
\affiliation{School of Physics, Huazhong University of Science and Technology, Wuhan, 430074, China}

\author{Taotao Qiu}
\email{qiutt@hust.edu.cn}
\thanks{Corresponding author.}
\affiliation{School of Physics, Huazhong University of Science and Technology, Wuhan, 430074, China}

\begin{abstract}
The Pulsar Timing Array (PTA) data of nano-Hertz gravitational waves released in 2023 implies that if such gravitational waves come from the scalar perturbation induction at the end of inflation, the accompanying primordial black holes (PBHs) will be over-produced, with the fraction exceeding the upper bound of unity. This is recognized as the ``overproduction problem'', which calls for nontrivial features in the early universe. In this paper, we try to check out whether a modified dispersion relation (MDR) of the primordial perturbations can be helpful for solving the problem. From the constraint on PTA data, we obtain a posterior distribution of the parameters of primordial perturbation, and find that the MDR model, where the $k^4$ term becomes important at later time, can give rise to a broken-power-law (BPL) power spectrum, which can bring the $f_{\rm PBH}=1$ line within $2\sigma$ level of the PTA constraint even in the absence of non-Gaussianity, indicating an avoidance of overproduction problem. However, to improve further to within $1\sigma$ still needs small negative non-Gaussianity, e.g. $f_{\rm nl}\simeq -1$. The mass distribution of the PBHs generated is also discussed.
\end{abstract}

\maketitle

\section{Introduction}

The Primordial Black Holes (PBHs) \cite{Zeldovich:1967lct, Hawking:1971ei, Carr:1974nx} are interesting in many research fields such as gravity, cosmology and astrophysics. Being generated from overdense spacetime curvature fluctuations in the very early universe, the PBHs can escape from the constraint of Chandrasekhar limit, and thus have a wider mass range than astrophysical black holes with fruitful properties. Not only is it possible for PBHs to act as $100\%$ dark matter, but also they can produce gravitational waves and/or Hawking radiation, which can be used as a probe of the early universe. See \cite{Khlopov:2008qy, Yuan:2021qgz, Villanueva-Domingo:2021spv, Oncins:2022ydg, Escriva:2022duf, Carr:2025kdk} for recent reviews and references therein.

It is well known that in order to generate PBH in the framework of inflation, a curvature perturbation with power spectrum enhanced to ${\mathcal O}(10^{-2})$ is required. According to the expression of the power spectrum, $P_\zeta\sim H^2/(\epsilon c_s)$, one of the interesting possibilities is to suppress the sound speed $c_s$ to $c_s\ll 1$ at the end of inflation. However, as has been pointed out in Ref. \cite{Cheung:2007st, Ballesteros:2018wlw}, the suppression of $c_s$ alone will lead to strong coupling problem. This problem can be avoided by introducing a higher-order term such as $k^4$ term, whereas the dispersion relation of the inflation is modified as  $\omega^2=c_s^2k^2+\alpha_M^2k^4$ \cite{Arkani-Hamed:2003juy,Arkani-Hamed:2003pdi,Ballesteros:2021fsp}. Such a modified dispersion relation (MDR) has been realized in the models of field theories \cite{Arkani-Hamed:2003juy,Arkani-Hamed:2003pdi} and modified gravity theories \cite{Qiu:2015aha, Qiu:2018nle}, while in the framework of these models,  the possibilities of generating PBHs have also been discussed \cite{Ballesteros:2018wlw, Ballesteros:2021fsp,Gorji:2021isn,Qiu:2022klm}.  

The latest observations of gravitational waves with the pulsar timing array (PTA) have been released almost simultaneously by NANOGrav 15yr \cite{NANOGrav:2023gor, NANOGrav:2023hvm}, PPTA \cite{Reardon:2023gzh, Zic:2023gta}, EPTA (with InPTA) \cite{EPTA:2023fyk, EPTA:2023gyr} and CPTA \cite{Xu:2023wog}. By observing a group of millisecond pulsars, they presented the signal correlations of Hellings-Downs angular pattern from $2\sigma$ to over $4\sigma$, which indicates the evidence for stochastic gravitational wave background (SGWB). Moreover, a strain spectrum of $A\sim 10^{-15}$ at a reference of $1\,\mathrm{yr}^{-1}$ (or nHz) has been found. Although sources from a population of supermassive black hole binary (SMBHB) are consistent, it shows that more exotic cosmological and astrophysical sources are more favored. As a precise example, the NANOGrav 15yr data prefers a power-law exponent of gravitational wave power spectral density $\gamma=3.2\pm0.6$ \cite{NANOGrav:2023gor, NANOGrav:2023hvm, NANOGrav:2023hfp} while the SMBHB gives rise to a steeper $\gamma=13/3$ \cite{Phinney:2001di}. For this reason, people are considering other mechanisms for this PTA signal, such as environment effects \cite{Ellis:2023dgf}, or other cosmological sources, such as primordial gravitational waves (pGWs) \cite{Grishchuk:1974ny, Starobinsky:1979ty}, scalar-induced gravitational waves (SIGWs) \cite{Tomita:1967wkp,Matarrese:1993zf,Matarrese:1997ay, Acquaviva:2002ud, Ananda:2006af, Domenech:2021ztg, Gangopadhyay:2023qjr}, first-order phase transition (FOPT) \cite{Witten:1984rs, Hogan:1986dsh}, topological defects \cite{Vilenkin:1984ib, Burden:1985md, Hindmarsh:1994re}, axion-like particles \cite{Guo:2023hyp}, Hawking radiation from PBHs \cite{Dong:2015yjs} and so on. See \cite{Ellis:2023oxs} for review about these sources. 

Among those candidates of PTA sources mentioned above, there is an interesting possibility that it could be gravitational waves induced by scalar primordial perturbations. It has been shown in Ref. \cite{Figueroa:2023zhu} with the model comparison Bayesian analysis that, the SIGW signal provides a better fit to the PTA data than the astrophysical counterpart. Moreover, it also provides a mechanism for PBH formation. But wait. What about the amount of the PBHs if we require the SIGW be suitable to explain the PTA data? In standard cosmology, both PBHs and its accompanying SIGWs are produced by the small-scale perturbations that are generated near the end of inflation. However, in Ref. \cite{Franciolini:2023pbf}, it was claimed that for single-field inflationary scenarios without non-Gaussianity, or even with positive non-Gaussianity, the amount of PBHs will be overproduced (with its fraction $f_{\rm PBH}\geq 1$) when the SIGW spectrum is required to explain the PTA data. The same results are obtained in \cite{Dandoy:2023jot} even for earlier NANOGrav 12.5yr observational data analysis. Such a tension, dubbed as the ``overproduction problem'', will thus place severe constraints on inflationary models which is responsible for PBH production. 

The requirement to overcome this issue places constraints on the inflation models, which requires them to possess certain features. Naively speaking, it may be helpful to reduce the population of the formed PBHs with either a large negative non-Gaussianity \cite{Franciolini:2023pbf, Wang:2023ost, Liu:2023ymk, DeLuca:2023tun, Firouzjahi:2023xke, Chang:2023aba, Pi:2024lsu, Inui:2024fgk} or a different window function/threshold value \cite{Inomata:2023zup}, see \cite{Unal:2023srk, Geller:2023shn, HosseiniMansoori:2023mqh, Balaji:2023ehk, Gorji:2023sil, Zhu:2023gmx, Liu:2023pau, Frosina:2023nxu, Bhaumik:2023wmw, Choudhury:2023hfm, Yi:2023npi, Choudhury:2023fwk, Liu:2023hpw, Choudhury:2023fjs, Choudhury:2024one, Wang:2024euw, Domenech:2024rks, Papanikolaou:2024fzf, Choudhury:2024dzw, Choudhury:2024kjj, Yogesh:2025hll, Kumar:2025jfi} for concrete models. Moreover, people have considered a constant $c_s$ different from unity during both inflation \cite{Choudhury:2023fwk,Choudhury:2023fjs} and the radiation-dominant era \cite{Balaji:2023ehk,Liu:2023hpw}. The former will affect the primordial power spectrum generated during inflation, while the latter will affect the transfer function in the calculation of the SIGW. However, the case where $c_s$ varies during the inflation era is less discussed. In this work, we'll explore whether this issue can be avoided in the case where the dispersion relation of the inflaton field gets modified, with a varying $c_s$ and a correction of $k^4$ term.

The rest of the paper is arranged as follows: In Sec. \ref{sec:pert} we calculate the primordial perturbations of general inflation model with modified dispersion relation, and obtain the curvature power spectrum. In Sec. \ref{sec:IGW} we discuss the gravitational waves induced by such a power spectrum, and constrain the parameters using the recent PTA data. In Sec. \ref{sec:PBH} we consider the PBHs generated at the end of inflation, and compare the fraction of PBHs to the data constraint to see whether the overproduction issue can be effectively avoided. Sec. \ref{sec:conclusion} contains our conclusions and discussions. 

\section{primordial perturbation with modified dispersion relation}
\label{sec:pert}
We consider the inflation model where the dispersion relation of the inflation perturbation is modified with a $k^4$ correction. While a typical model can be realized in  \cite{Arkani-Hamed:2003pdi, Arkani-Hamed:2003juy, Qiu:2015aha, Qiu:2018nle,Ballesteros:2021fsp,Gorji:2021isn,Qiu:2022klm}  (see example of model building in Appendix \ref{appdx}), in general case without model specification, the Mukhanov-Sasaki equation of the curvature perturbation can be written as:
\begin{equation}
\label{zetabar}
{u_k''} {\rm{ + }}\left[ {c_s^2{k^2} + {{\alpha}_M^2}{k^4}{\tau ^2} - \frac{{{\theta ^2} - 1/4}}{{{\tau ^2}}}} \right]{u _k} = 0~.
\end{equation}
Here, $c_s$ stands for the sound speed of the perturbation, $\tau$ stands for the conformal time and $\alpha_M$ is the coefficient of the $k^4$ term. The variable $u_k$ is a redefinition of the curvature perturbation $\zeta_k$, with the relation
\begin{equation}%
\label{zeta}
u_k =z\zeta_k~,~~~ z \propto a\left( \tau  \right)~.
\end{equation}
Note that for $c_s=1$ and $\alpha_M=0$, Eq. \eqref{zetabar} will go to the trivial case for pure slow-roll inflation \footnote{Here we turn down the effects from $\theta$ as well, which is not the focus on this paper.} with power spectrum $P_\zeta\sim 10^{-9}$, thus can hardly generate PBHs or SIGWs because the scalar perturbations are too small. Now we we consider the varying sound speed as:
\begin{equation}%
\label{soundspeed}
c_s  \left\{ {\begin{array}{*{20}{r}}
\simeq 1~,&\tau<\tau_\ast\\
\ll 1~,&\tau>\tau_\ast
\end{array}} \right.
\end{equation}
where $\tau_\ast$ is some pivot time \cite{Gorji:2021isn, Qiu:2022klm}. We assume for simplicity that in regions both before and after $\tau_\ast$ the inflation is slow-rolling, therefore, it is also convenient to dub the two regions as ``slow-roll I phase'' and ``slow-roll II phase'' respectively. Then Eq. \eqref{zetabar} turns out to be
\begin{align}\label{sreq}
{u^{\rm{srI}}_k}'' +\left[ {c_s^2{k^2}  - \frac{{{\theta ^2} - 1/4}}{{{\tau ^2}}}} \right]{u^{\rm{srI}}_k} = 0~~~
\text{for slow-roll I phase}~, \\
\label{steq}
{u^{\rm{srII}}_k}''+\left[ {{{\alpha}_M^2}{k^4}{\tau ^2} - \frac{{{\theta ^2} - 1/4}}{{{\tau ^2}}}} \right]{u^{\rm{srII}}_k} = 0~~~
\text{for slow-roll II phase}~.
\end{align}

We note that Eqs.~\eqref{sreq} and \eqref{steq} are obtained from Eq. \eqref{zetabar} with Eq. \eqref{soundspeed} applied. The reduction of the dispersion relation is due to the behavior of the sound speed before and after the pivot time. For slow-roll I phase, the $k^4$ term is subdominant in superhorizon region, and when we consider modes with not too large $k$, satisfying the condition $-\alpha_Mk^2\tau^2/2<-c_sk\tau$, it cannot be dominant in subhorizon region as well \cite{Qiu:2022klm}. Therefore the equation remains the same as that for normal slow-roll case. For slow-roll II phase, the $k^2$ term is subdominant because of the vanishing $c_s$. The concrete model in Appendix \ref{appdx} shows explicitly the validation of such a reduction.

It is straightforward to get the solution in both phases, which has been done in \cite{Gorji:2021isn, Qiu:2022klm}: 
\begin{eqnarray}%
\label{srso}
{u^{\rm{srI}}_k}(k,\tau)&=&\sqrt{-\tau}\left[C^{\rm{srI}}_{1}H^{(1)}_{3/2}(-c_sk\tau)+C^{\rm{srI}}_{2}H^{(2)}_{3/2}(-c_sk\tau)\right]~,\\
\label{stso}
{u^{\rm{srII}}_k}(k,\tau)&=&\sqrt{-\tau}\left[C^{\rm{srII}}_{1}H^{(1)}_{3/4}\left(\frac{\alpha_M k^2\tau^2}{2}\right)+C^{\rm{srII}}_{2} H^{(2)}_{3/4}\left(\frac{\alpha_M k^2\tau^2}{2}\right)\right]~,
\end{eqnarray}
where $C^{\rm{srI}}_{1}$, $C^{\rm{srI}}_{2}$, $C^{\rm{srII}}_{1}$, $C^{\rm{srII}}_{2}$ are constant coefficients, while $H^{(1)}$ and $H^{(2)}$ are the first and the second Hankel functions.

The coefficients of the slow-roll I phase solution can be determined by the initial condition of inflation, namely Bunch-Davies vacuum solution. This gives the coefficients as:
\begin{equation}
   C^{\rm{srI}}_{1}=\frac{\sqrt{\pi}}{2}~,~~~C^{\rm{srI}}_{2}=0~.
\end{equation}  
On the other hand, in order to determine the coefficients of the slow-roll II solution, we impose the junction relations at the pivot time $\tau_\ast$ proposed in \cite{Gorji:2021isn, Qiu:2022klm}:
\begin{align}
\label{junction}
u^{\rm{srI}}_{k}|_{\tau=\tau_\ast}=u^{\rm{srII}}_{k}|_{\tau=\tau_\ast}~,~~~{u^{\rm{srI}}_{k}}'|_{\tau=\tau_\ast}={u^{\rm{srII}}_{k}}'|_{\tau=\tau_\ast}~.
\end{align}
This will give rise to:
\begin{eqnarray}%
\label{coefficient}
\left[ {\begin{array}{*{20}{c}}{{C^{\rm{srII}}_1}}\\{{C^{\rm{srII}}_2}}\end{array}} \right] &=&{\frac{{{\rm{\pi }}\alpha_M {k^2}}}{{8{\rm{i}}}}{C^{\rm{srI}}_1}\tau_\ast^2}\nonumber\\
&&\times
\left[ {\begin{array}{*{20}{c}}
{{\rm{ + }}{c_s}{{\left( { - \alpha_M k \tau _\ast} \right)}^{ - 1}}H_{3/4}^{(2)}\left( {\frac{{\alpha_M {k^2}}}{2}\tau_\ast^2} \right)H_{1/2}^{(1)}\left( { - c_sk{\tau _\ast}} \right) - H_{ - 1/4}^{(2)}\left( {\frac{{\alpha_M {k^2}}}{2}\tau _\ast^2} \right)H_{3/2}^{(1)}\left( { - c_sk{\tau _\ast}} \right)}\\
{ - {c_s}{\left(  - \alpha_M k \tau_\ast \right)}^{ - 1}}H_{3/4}^{(1)}\left( {\frac{{\alpha_M {k^2}}}{2}\tau _\ast^2} \right)H_{1/2}^{(1)}\left( { - c_sk{\tau _\ast}} \right) + H_{ - 1/4}^{(2)}\left( {\frac{{\alpha_M {k^2}}}{2}\tau _\ast^2} \right)H_{3/2}^{(1)}\left( { - c_sk{\tau _\ast}} \right)\end{array}} \right]~.
\end{eqnarray}

\begin{figure}
    \centering
    \includegraphics[width=0.8\linewidth]{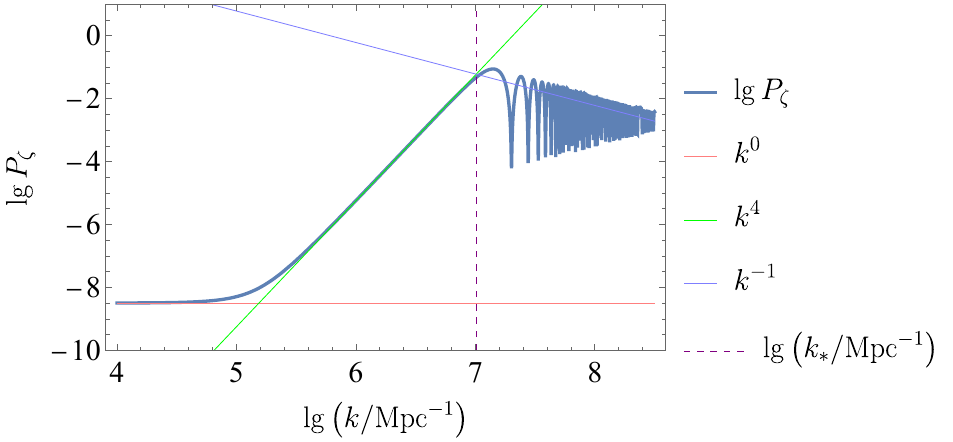}
    \caption{The power spectrum of curvature perturbation obtained from Eq. \eqref{zetabar} with respect to $k$ (in logarithm). Three straight lines mimicking the $P_\zeta-k$ relations are also shown for comparison: red line is for $P_\zeta\sim k^0$, green is for $P_\zeta\sim k^4$, while blue is for $P_\zeta\sim k^{-1}$. The dashed line denotes the pivot scale $k_\ast$ corresponding to the peak of the power spectrum. With the power spectrum in large scales around $2.0\times10^{-9}$, ${\lg|\tau _*|}$ and $\lg \alpha_M$ are set to be $-4.95$ and $-3.86$ to make the BPL parameters $\lg(k_*/\mathrm{Mpc^{-1}})$ and $\lg A$ to be $7.0$ and $-1.4$. }
    \label{fig:curvpert}
\end{figure}

We obtain the power spectrum of the curvature perturbations at their values at the end of inflation, namely $\tau\rightarrow 0$. In this limit, solution \eqref{stso} will have the following approximate behavior:
\begin{equation}
    \label{stsolutionend}
    u^{\rm{srII}}_{k} = {2^{3/2}}{\rm{i}}{{\rm{\pi }}^{ - 1}}{\rm{\Gamma }}\left( {3/4} \right){\alpha_M ^{ - 3/4}}|{{C^{\rm{srII}}_2} - {C^{\rm{srII}}_1}}|{k^{ - 3/2}}{\left( { - \tau } \right)^{ - 1}}~.
\end{equation}
Due to the expressions \eqref{coefficient}, the behavior of Eq. \eqref{stsolutionend} will be different in different scales, labeled with $k$. We split the solution \eqref{stsolutionend} into three branches corresponding to three types of scales, according to the magnitude of the terms $\alpha_M k^2\tau^2/2$ and $-c_sk\tau$ relative to unity. 
\footnote{For evolutions of $u_k$ before the pivot scale $\tau_\ast$, we always have $\alpha_M {k^2}\tau^2/2<-c_sk{\tau}$, see Fig. 1 (right panel) in \cite{Qiu:2022klm}.} In large scales where $\alpha_M {k^2}\tau^2/2<-c_sk{\tau}<1$, the solution becomes
\begin{equation}
    \label{large}
 u^{\rm{srII}}_{k} =  - 2{\rm{i}}{{\rm{\pi }}^{ - 2}}{\rm{\Gamma }}\left( {3/4} \right){\rm{\Gamma }}\left( {{\rm{1}}/{\rm{4}}} \right){\rm{\Gamma }}\left( {{\rm{3}}/{\rm{2}}} \right)c_s^{-3/2}{C^{\rm{srI}}_1}{k^{ - 3/2}}{\left( { - \tau } \right)^{ - 1}}~.
\end{equation}
In medium scales where $\alpha_M {k^2}\tau^2/2 < 1 <  - c_sk{\tau}$, it becomes
 \begin{equation}
    \label{medium}
u^{\rm{srII}}_{k} = {\rm{i}}\frac{4}{3}{2^{ - 3/2}}{{\rm{\pi }}^{ - 1/2}}c_s^{1/2}\tau _\ast^2{{\rm{e}}^{ -{{\rm{i}} c_sk{\tau _\ast}}}}{C^{\rm{srI}}_1}{k^{1/2}}{\left( { - \tau } \right)^{ - 1}}~.
\end{equation}
While in small scales where $1 < \alpha_M {k^2}\tau^2/2 <  - c_sk{\tau}$, it becomes
\begin{equation}
   \label{small}
   u^{\rm{srII}}_{k}=\sqrt 2 {{\rm{\pi }}^{ - 1}}{\rm{i}}{{\rm{e}}^{-{\rm{i}} c_sk{\tau _\ast}}}{\rm{\Gamma }}\left( {3/4} \right){\alpha_M ^{ - 5/4}}c_s^{1/2}{\left( { - {\tau _\ast}} \right)^{ - 1/2}}{C^{\rm{srI}}_1}{k^{ - 2}}{\left( { - \tau } \right)^{ - 1}}~.
\end{equation}
The power spectrum of the curvature perturbation is defined as follows:
\begin{equation}
    {P_\zeta}(k) \equiv \frac{k^3}{2\pi^2}\left|\frac{u_k}{z}\right|^2~.
\end{equation}
Thus from Eqs. \eqref{large}, \eqref{medium} and \eqref{small} one finds that
\begin{equation}
\label{spectrum}
    {P_{\zeta}}(k) \propto \left\{ {\begin{array}{*{20}{r}}
k^0 & \text{for large scale~,}\\
k^4 & \text{for medium scale~,}\\
k^{-1} & \text{for small scale~.} 
\end{array}} \right.
\end{equation}
We numerically calculate the equation \eqref{zetabar} and plot the power spectrum in Fig. \ref{fig:curvpert}. The figure show nice consistency with the analytical result \eqref{spectrum}, however, there also exists some oscillations on the small scales. This is due to the fact that for very large $k$ modes which does not exit the horizon before the pivot scale $\tau_\ast$, the subhorizon effect will become robust. However, it can be smoothed to mimic the power-law form. Since in this work we're mainly focusing on the medium and small scales which are responsible for SIGW and PBH generation, it is useful to parametrize the power spectrum within these scales into a broken-power-law (BPL) form:
\begin{equation}
\label{BPL}
    {P_{\zeta}}\left( k \right) = \left\{ {\begin{array}{*{20}{c}}
{A{{\left( {\frac{k}{{{k_ * }}}} \right)}^4}}&{ k < {k_ * }}~,\\
{A{{\left( {\frac{k}{{{k_ * }}}} \right)}^{ - 1}}}&{k > {k_ * }}~,
\end{array}} \right.
\end{equation}
where $k_\ast$ is pivot scale between medium and small scales, and $A$ is the amplitude of power spectrum at the pivot scale.

We now make explicit how the BPL parameters $\{A,k_\ast\}$ that will be constrained by the PTA data later are related to the parameters $\alpha_M$ and $\tau_\ast$ of the MDR equation \eqref{zetabar}, so that the posterior obtained below can be translated back onto the parameters of the MDR model. In the present framework the sound speed is {\it not} a continuously free parameter: it takes its two limiting values, $c_s=1$ for $\tau<\tau_\ast$ (slow-roll I) and $c_s\to0$ for $\tau>\tau_\ast$ (slow-roll II), as in Eq.~\eqref{soundspeed}; in the slow-roll II phase the $c_s^2k^2$ term is therefore dropped from the mode equation, so that only the $k^4$ correction remains, cf.\ Eq.~\eqref{steq}. The medium-scale branch of Eq.~\eqref{spectrum} behaves as $P_\zeta\propto k^4$ and follows from Eq.~\eqref{medium}, while the small-scale branch $P_\zeta\propto k^{-1}$ follows from Eq.~\eqref{small}. The pivot scale $k_\ast$ of the broken-power-law spectrum \eqref{BPL} is located at the crossing of these two power laws. Using Eqs.~\eqref{medium}, \eqref{small} and the definition \eqref{spectrum}, the crossing condition $P_\zeta^{(m)}(k_\ast)=P_\zeta^{(s)}(k_\ast)$ yields
\begin{equation}
\label{eq:MDR-BPL-kstar}
k_\ast \;\simeq\; \frac{c_0}{\sqrt{\alpha_M}\,|\tau_\ast|} \ ,\qquad
c_0\;\equiv\;\left[\frac{9\,\Gamma^2(3/4)}{\pi}\right]^{1/5}\simeq 1.34 \ ,
\end{equation}
i.e.\ for a given pivot time $\tau_\ast$ the coefficient $\alpha_M$ is fixed by the pivot scale through the combination $\alpha_M k_\ast^2\tau_\ast^2\simeq c_0^2\simeq1.79$. As a cross-check, the numerical realization of Fig.~\ref{fig:curvpert} ($\lg|\tau_\ast|=-4.95$, $\lg\alpha_M=-3.86$, quoted as $\lg(k_\ast/\mathrm{Mpc}^{-1})=7.0$) gives $\alpha_M k_\ast^2\tau_\ast^2\simeq10^{0.24}\simeq1.74$, in agreement at the few-percent level. Next, the overall normalization of $P_\zeta$ is fixed by the {\it large}-scale branch: identifying the plateau of Eq.~\eqref{large} (with $c_s=1$) with the CMB amplitude $\mathcal{P}_\zeta^{\mathrm{CMB}}\simeq2\times10^{-9}$ adopted in Fig.~\ref{fig:curvpert} determines the combination $H^2/\beta^2$ that multiplies the medium-scale branch (here $\beta$ is the coefficient in $z=\beta a$, normalized by the standard slow-roll relation $z^2\simeq2\epsilon M_p^2a^2$, see Appendix~\ref{appdx}). Substituting Eq.~\eqref{eq:MDR-BPL-kstar} into the amplitude of the medium-scale branch and imposing $A=P_\zeta(k_\ast)$ at the crossing, the peak value reads
\begin{equation}
\label{eq:MDR-BPL-A}
A \;\simeq\; \frac{c_0^4}{9}\;\mathcal{P}_\zeta^{\mathrm{CMB}}\;\frac{1}{\alpha_M^2}
\;\simeq\; 0.36\left(\frac{\mathcal{P}_\zeta^{\mathrm{CMB}}}{2\times10^{-9}}\right)\frac{1}{\alpha_M^2} \ .
\end{equation}

Two remarks are in order:\\
\textit{(i) Log-linear dictionary and inverse.} Equations \eqref{eq:MDR-BPL-kstar} and \eqref{eq:MDR-BPL-A} are manifestly log-linear:
\begin{align}
\label{eq:MDR-BPL-log}
\lg A &=\lg\mathcal{P}_\zeta^{\mathrm{CMB}}+\lg\left(c_0^4/9\right)-2\lg\alpha_M\ ,\\
\lg\left(\frac{k_\ast}{\mathrm{Mpc}^{-1}}\right) &=\lg c_0-\tfrac12\lg\alpha_M-\lg\left(\frac{|\tau_\ast|}{\mathrm{Mpc}}\right)\ .
\end{align}
Therefore, given a point $(A,k_\ast)$ of the PTA posterior, the MDR parameters are recovered through
\begin{equation}
\label{eq:MDR-BPL-inv}
\lg\alpha_M=-\tfrac12\left[\lg A-\lg\mathcal{P}_\zeta^{\mathrm{CMB}}-\lg\left(c_0^4/9\right)\right]\ ,\qquad
\lg\frac{|\tau_\ast|}{\mathrm{Mpc}}=\lg c_0-\tfrac12\lg\alpha_M-\lg\frac{k_\ast}{\mathrm{Mpc}^{-1}}\ .
\end{equation}
In particular, $\lg\alpha_M$ follows from $\lg A$ by a pure rescaling with slope $-1/2$, whereas $\lg(|\tau_\ast|/\mathrm{Mpc})$ is the linear combination $\tfrac14\lg A-\lg(k_\ast/\mathrm{Mpc}^{-1})+\mathrm{const}$. Evaluating Eqs.~\eqref{eq:MDR-BPL-log} at the benchmark values of Fig.~\ref{fig:curvpert} ($\lg|\tau_\ast|=-4.95$, $\lg\alpha_M=-3.86$) reproduces $\lg(k_\ast/\mathrm{Mpc}^{-1})\simeq7.0$ and gives $\lg A\simeq-1.43$.\\
\textit{(ii) Scope.} The dictionary above applies to the strictly two-stage model adopted in this paper, in which the sound speed takes the limiting values $c_s=1$ in the slow-roll I phase and $c_s\to0$ in the slow-roll II phase, as assumed in Eq.~\eqref{soundspeed}: $\alpha_M$ and $\tau_\ast$ are then the only MDR parameters left, and the two PTA-determined quantities $(A,k_\ast)$ constrain them one-to-one through Eq.~\eqref{eq:MDR-BPL-inv}. Since the likelihood analysis of Sec.~\ref{sec:IGW} is performed at the level of the BPL parameters (because of the $\mathcal{O}(10\%)$ systematic uncertainty mentioned above), the translation of the benchmark posterior points through Eq.~\eqref{eq:MDR-BPL-inv} directly returns the typical values of the MDR parameters.

\section{gravitational waves induced from scalar perturbation}
\label{sec:IGW}
It is well-known that the scalar and tensor perturbations decouple from each other in linear level. However, when going into non-linear level, the coupling will appear, and the scalar perturbation can be transferred into the tensor one via the scalar-tensor interaction, while the latter consists of the gravitational waves. Although it has been suppressed during most part of the inflation era, at the end of inflation where the scalar perturbation gets enlarged, such an interaction becomes important, and the induced tensor perturbation can act as a signal to be observed by the current gravitational wave observations. Here we consider the induced gravitational waves generated by the scalar perturbations given in \eqref{spectrum}.

The energy density fraction spectrum of the present-day SIGW is calculated in
\cite{PhysRevD.75.123518,Kohri:2018awv} (see also \cite{Domenech:2021ztg} for a review). For scalar-induced case, the equation of motion for tensor perturbations can be written as:
\begin{equation}
\label{eomhk}
    h''_{\bf k}(\tau)+2aHh'_{\bf k}(\tau)+k^2h_{\bf k}(\tau)=4S_{\bf k}(\tau)~.
\end{equation}
Here $S_{\bf k}(\tau)$ is the source term from scalar perturbation. Since the SIGW is generated in the Radiation Domination (RD) era as the relevant mode reenters the horizon \cite{Baumann:2007zm,Kohri:2018awv}, where we assume that the single inflation field $\phi$ decays out, one will get an EoM the same as in normal GR: 
\begin{equation}
\label{source}
    S_{\bf k}(\tau)=\int \frac{{{d^3}}q}{(2\pi)^{3/2}}e_{ij}({\bf k})q_iq_j\left[2\Phi_{\bf q}\Phi_{\bf k-q}+\frac{4}{3(1+w)}\left(\Phi_{\bf q}+\frac{\Phi_{\bf q}^\prime}{\cal H}\right)\left(\Phi_{\bf k-q}+\frac{\Phi_{\bf k-q}^\prime}{\cal H}\right) \right]~,
\end{equation}
where $\Phi_{\bf k}$ is the gravitational potential, and $e_{ij}({\bf k})$ is the polarization tensor. Thus, the solution of Eq. \eqref{eomhk}  takes the normal form as \cite{Domenech:2021ztg}:
\begin{equation}
\label{solhk}
    h_{\bf k}(\tau)=\frac{4}{a}\int^\tau_{\tau_0} {{d}}\tau' G_{\bf k}(\tau,\tau')a(\tau')S_{\bf k}(\tau')~,
\end{equation}
where $G_{\bf k}(\tau,\tau')$ is the Green function.

In the RD era, the gravitational potential satisfies the equation
\begin{equation}
    \Phi''_{\bf k}(\tau)+\frac{6(1+w)}{(1+3w)\tau}\Phi'_{\bf k}+wk^2\Phi_{\bf k}(\tau)=0~,
\end{equation}
where one has $w=1/3$. This gives rise to the solution: 
\begin{eqnarray}
    \Phi_{\bf k}(\tau)&=&\Phi_0({\bf k}){\cal T}(k,\tau)~,\\
    \label{transfunc}
    {\cal T}(k,\tau)&=&3\frac{{\sin (k\tau /\sqrt 3) - (k\tau /\sqrt 3)\cos (k\tau /\sqrt 3)}}{(k\tau /\sqrt 3)^3}~.
\end{eqnarray}
The coefficient $\Phi_0({\bf k})$ is the initial condition of the solution for this era, which will be connected to the perturbations during inflation. As the Newtonian potential, it relates to the curvature perturbation $\zeta$ as 
\begin{equation}
\label{phizeta}
    \Phi_0({\bf k})\simeq\frac{\epsilon}{1+\epsilon}\zeta_{\bf k}=\frac{3+3w}{5+3w}\zeta_{\bf k}~.
\end{equation}

The tensor spectrum is defined as:
\begin{equation}
\label{tensorspectrum}
    \langle h_{\bf k}(\tau)h_{{\bf k}'}(\tau)\rangle=\delta^3({\bf k}+{\bf k}')\frac{2\pi^2}{k^3}P_h(k,\tau)~.
\end{equation} 
Meanwhile, the energy density of gravitational waves is defined as:
\begin{equation}
    \rho_{\rm GW}(k,\tau)\equiv\frac{d\rho_{\rm GW}}{d\ln k}=\frac{M_p^2}{16a^2}\frac{d}{d\ln k}\langle h_{ij,{\bf k}}h_{ij,{\bf k}}\rangle~,
\end{equation} 
note that here $h_{ij}$ is Fourier conjugate of the $h_{\bf k}(\tau)$ in Eq. \eqref{tensorspectrum}, and $h_{ij,{\bf k}}\equiv\partial_{\bf k}h_{ij}$. Therefore, the energy density fraction of the gravitational waves is written as:
\begin{equation}
    \Omega_{\rm{GW}}(k,\tau)\equiv\frac{\rho_{\rm{GW}}(k,\tau)}{\rho_{\rm{tot}}(\tau)}=\frac{1}{24}\left(\frac{k}{a(\tau)H(\tau)}\right)^2P_h(k,\tau)~.
\end{equation}
At the horizon-reentrance time $\tau_k$ when $k=a(\tau_k)H(\tau_k)$, one has $\Omega^{\langle k\rangle}_{\rm{GW}}(k)=P^{\langle k\rangle}_h(k)/24$, where the subscript $\langle k\rangle$ denotes the values at reentrance time point $\tau_k$. On the other hand, for gravitational waves generated during radiation-dominated era, $\rho_{\rm{GW}}(k,\tau)=\Omega_{\rm{GW}}(k,\tau)\rho_{\rm{tot}}(\tau)$ evolves as $a^{-4}$ \cite{Kohri:2018awv, Domenech:2020xin}. Thus one has:
\begin{equation}
    \frac{\Omega^{\langle 0\rangle}_{\rm{GW}}(k)}{\Omega^{\langle k\rangle}_{\rm{GW}}(k)}=a^4(\tau_k)\left(\frac{\rho^{\langle k\rangle}_{\rm{tot}}}{\rho^{\langle 0\rangle}_{\rm{tot}}}\right)=\left(\frac{g^{\langle k\rangle}_\ast}{g^{\langle 0\rangle}_\ast}\right)\left(\frac{g^{\langle k\rangle}_{\ast S}}{g^{\langle 0\rangle}_{\ast S}}\right)^{-\frac{4}{3}}\Omega^{\langle 0\rangle}_r~,
\end{equation}
where the subscript $\langle 0\rangle$ denotes the values at the current time, and $\Omega^{\langle 0\rangle}_{r}\equiv\rho^{\langle 0\rangle}_r/\rho^{\langle 0\rangle}_{\rm{tot}}\simeq 4.3\times10^{-5}h^{-2}$. Here we have made use of the fact that in radiation-dominated era $\rho^{\langle k\rangle}_{\rm{tot}}=\rho^{\langle k\rangle}_{r}$, $\rho_{r}(\tau)=(\pi^2/30)g_\ast T^4$ and the entropy $s\sim g_{\ast S}T^3\sim a^{-3}$ (adiabatic condition). Then 
the current energy density fraction, which is to be compared to observations, is
\begin{equation}
    \Omega^{\langle 0\rangle}_{\rm{GW}}(k)= \frac{\Omega^{\langle 0\rangle}_r}{{24}}\left(\frac{g^{\langle k\rangle}_\ast}{g^{\langle 0\rangle}_\ast}\right)\left(\frac{g^{\langle k\rangle}_{\ast S}}{g^{\langle 0\rangle}_{\ast S}}\right)^{-\frac{4}{3}}P^{\langle k\rangle}_h(k)~.
\end{equation}

Using Eqs. \eqref{solhk} and \eqref{phizeta} and after tedious calculation, one gets the spectrum of tensor perturbation at the horizon-reentrance as 
\begin{equation}
    P^{\langle k\rangle}_h(k) = 4\int_1^\infty  {{d}}t\int_0^1 {{d}}s{{\left[ {\frac{{\left( {{t^2} - 1} \right)\left( {1 - {s^2}} \right)}}{{{t^2} - {s^2}}}} \right]}^2}{I^2}\left( {t,s} \right){P_\zeta }\left( {k\frac{{t - s}}{2}} \right){P_\zeta }\left( {k\frac{{t + s}}{2}} \right)~,
\end{equation}
while the transfer function $I(t,s)$ is given by:
\begin{equation}
        {I^2}(t,s)  = 288\frac{(s^2 + t^2 - 6)^2}{(t^2 - s^2)^6}\bigg[ \frac{\pi^2}{4}(s^2 + t^2 - 6)^2\Theta(t - \sqrt{3}) + \left( {t^2 - s^2 - \frac{1}{2}(s^2 + t^2 - 6)\ln \left| \frac{t^2 - 3}{3 - s^2} \right|} \right)^2 \bigg]~,
        \label{I}
\end{equation}
where $t\equiv (|{\bf k}-{\bf q}|+q)/k$, $s\equiv(|{\bf k}-{\bf q}|-q)/k$. 

\begin{figure}
    \centering
    \includegraphics[width=0.75\linewidth]{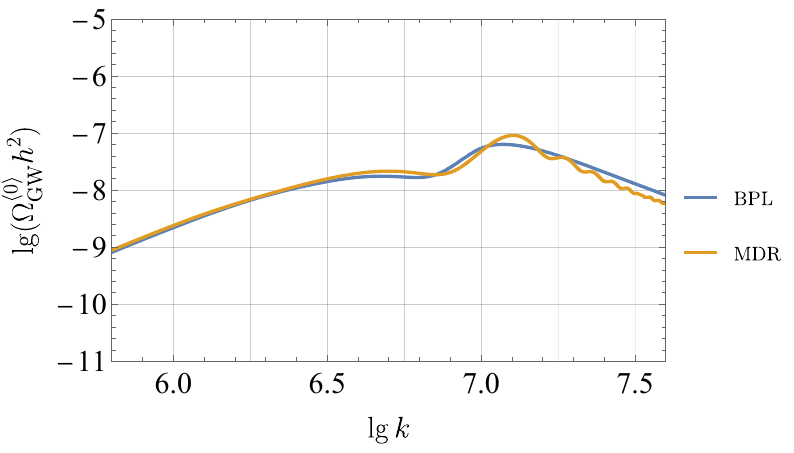}
    \caption{The numerical calculations of induced gravitational waves where the power spectrum is given by MDR model (orange) and BPL parametriztion (blue) respectively. The parameters of MDR and BPL models are the same as in Fig. \ref{fig:curvpert}. }
    \label{fig:IGWcompare}
\end{figure}

\begin{figure}
    \centering    \includegraphics[width=0.6\linewidth]{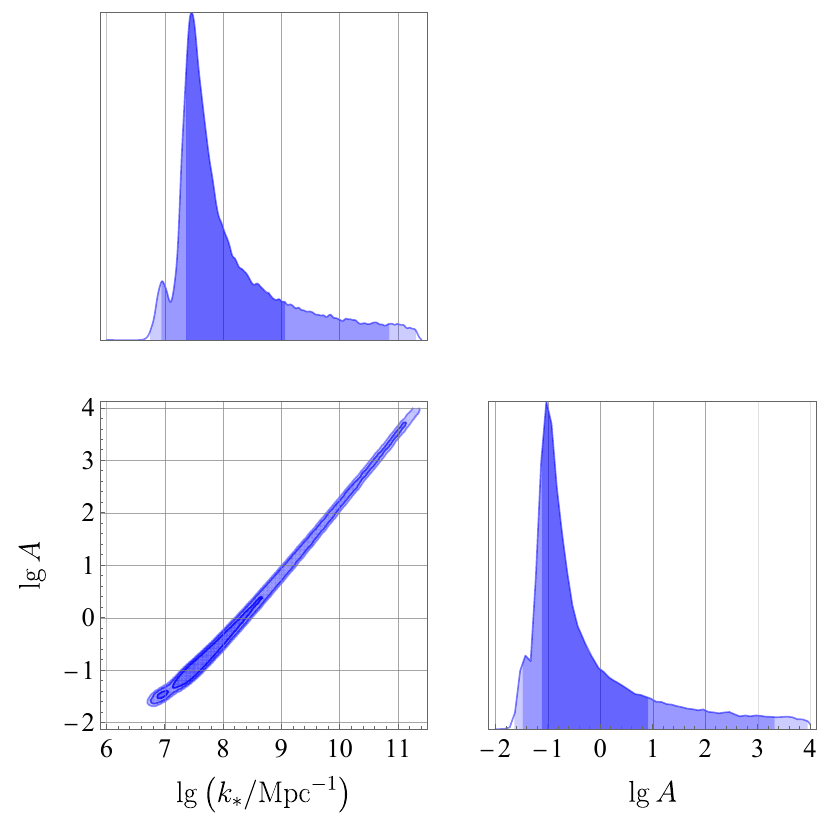}
    \caption{Posterior of the parameters of the BPL model, where NANOGrav 15-yr data was used. The contours of the 2D posterior plot from dark to light corresponds to the 1$\sigma$, 2$\sigma$ and 3$\sigma$ confidence levels, respectively. Priors of the parameters are set as $\lg A\in[-3,4]$ and $\lg(k_\ast/\mathrm{Mpc}^{-1})\in[5.8,12.8]$. The plotted range is chosen so as to cover the full region where the posterior is non-negligible; the zoomed views of the parameter region relevant for the PBH analysis are shown in Figs. 5–6.
}
    \label{fig:SIGW}
\end{figure}

Since the tensor spectrum contains complicated integrations, it is then difficult to make predictions analytically, so we refer to numerical solutions as well. First of all, in Fig. \ref{fig:IGWcompare} we calculated numerically $\Omega^{\langle 0\rangle}_{\rm GW}(k)$ (in logarithm) using power spectrum from MDR model (in orange) and the BPL parametrization (in blue) respectively, finding they are consistent with each other as expected. Therefore it is appropriate and useful to calculate the posterior distributions using the parametrized form instead of the model, which is much more complicated. In Fig. \ref{fig:SIGW} we plot the posterior distributions for the amplitude of curvature perturbation as well as the scales. In this plot, we use NANOGrav 15-yr data as well as the BPL parametrization as shown in Eq. \eqref{BPL}. The contour in Fig. \ref{fig:SIGW} is not too different from previous works with BPL curvature spectrum, which is because the result actually has weak dependence on power-law indices of curvature power spectrum, except for requiring a blue-tilt by the PTA data. Moreover, as is shown in e.g. \cite{Cai:2018dig}, the non-Gaussian property of the curvature perturbations will not alter the result too much, either. Actually, the quadratic order non-Gaussian estimator $f_\mathrm{nl}$ is restricted to ${\cal{O}}(1)$ in the following text, see more discussions on larger $f_\mathrm{nl}$'s in e.g. \cite{Zeng:2025cer}.

As a supplementary statement, the posterior distributions presented in this work are obtained with the public code \texttt{PTArcade}~\cite{Mitridate:2023oar}, a Bayesian analysis framework for PTA data in which the likelihood and the NANOGrav 15-yr data set adopted here are already implemented. The convergence is monitored with the Gelman--Rubin diagnostic, which gives $R=1.00017$ for $\lg A$ and $R=1.00018$ for $\lg(k_\ast/\mathrm{Mpc}^{-1})$, i.e.\ $|R-1|\lesssim2\times10^{-4}$, far below the conventional threshold $|R-1|<0.01$; the corresponding total effective sample sizes are $22190$ and $22097$, respectively.

The apparent bimodality of the one-dimensional marginals of $\lg(k_\ast/\mathrm{Mpc}^{-1})$ and $\lg A$ also deserves a comment. In Fig.~\ref{fig:SIGW}, we have fixed the power-law index of $k<k_\ast$ and $k>k_\ast$ sections to be $4$ and $-1$. However, if we set them free, 
we recover the single-peaked posterior obtained in Ref.~\cite{Franciolini:2023pbf}. This indicates that the double-peaked structure might be induced by the restricted form of the template adopted here.

\begin{figure}
    \centering
    \includegraphics[width=0.6\linewidth]{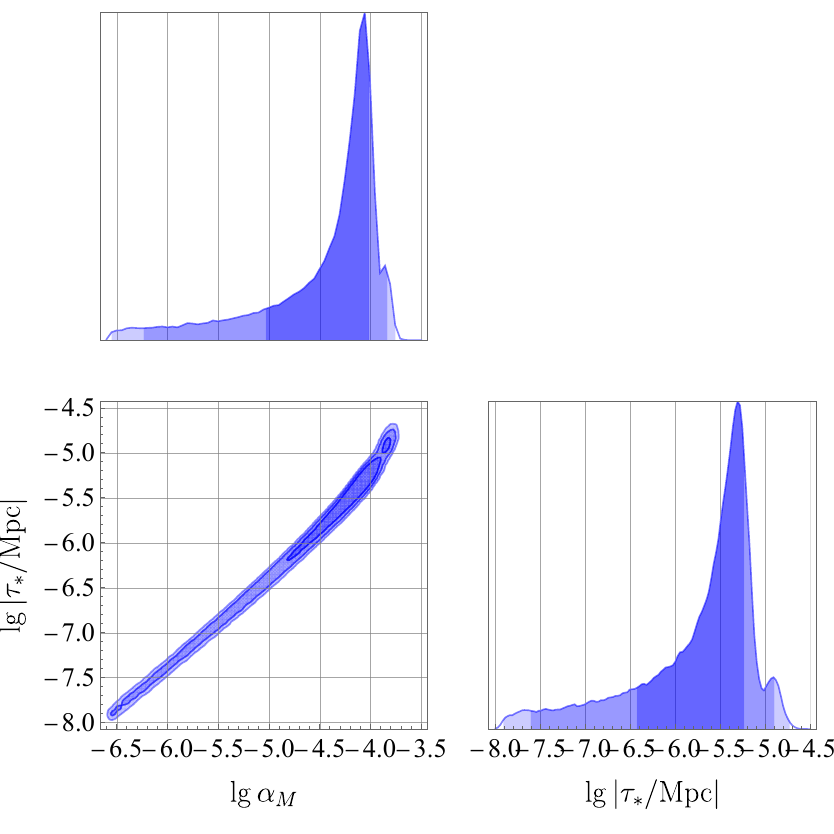}
    \caption{Posterior of the parameters of the MDR model translated from the BPL parameters through Eq. \eqref{eq:MDR-BPL-inv}, The contours of the 2D posterior plot from dark to light correspond to the 1$\sigma$, 2$\sigma$ and 3$\sigma$ confidence levels, respectively. 
    }
    \label{fig:posteriorMDR}
\end{figure}

Moreover, using the inverse relations \eqref{eq:MDR-BPL-inv} we can further translate the BPL posterior of Fig.~\ref{fig:SIGW} into the parameter space of the MDR model. The resulting one-dimensional marginal posterior distributions of $\lg\alpha_M$ and of $\lg(|\tau_\ast|/\mathrm{Mpc})$ are displayed in Fig.~\ref{fig:posteriorMDR}. Since the dictionary \eqref{eq:MDR-BPL-inv} is linear in the logarithms, these marginals inherit the shape of the corresponding marginals of $\lg A$ and $\lg(k_\ast/\mathrm{Mpc}^{-1})$, so that the two MDR parameters are strongly correlated along this direction.

\section{PBH production and overproduction avoidance}
\label{sec:PBH}
In this section we analyze the formation of PBHs and see whether the overproduction problem will be avoided in our model. To do this, we make use of the ``compaction function'' approach, which was introduced in \cite{Shibata:1999zs} and has been widely discussed in \cite{Harada:2015yda, Yoo:2018kvb, Kawasaki:2019mbl, Biagetti:2021eep, Kitajima:2021fpq, Young:2022phe, Escriva:2022pnz, Ferrante:2022mui, Gow:2022jfb}.  In this approach, one defines the compaction function as:
\begin{equation}
    {\cal{C}}(\boldsymbol{x},r)=2\frac{\delta M(\boldsymbol{x},r)}{R(r)}~,
\end{equation}
which describes the mass excess within a sphere of areal radius $R$ centered on spatial coordinate $\boldsymbol{x}$. The Misner-Sharp mass is defined as:
\begin{equation}
    M=\int 4\pi R^2\rho {{d}}R=\int 4\pi R^2R_{,r}\rho {{d}}r~,
\end{equation} 
where the subscript ``$,r$'' denotes derivative with respect to $r$. Therefore, the mass excess turns out to be:
\begin{eqnarray}
\label{deltaM}
    \delta M&=&4\pi a^3\rho_0\int\delta r^2e^{-3\zeta}(1-r\zeta_{,r}){{d}}r~,\nonumber\\
    &=&\frac{3(1+w)}{2(5+3w)}ar^2e^{-\zeta}\zeta_{,r}(2-r\zeta_{,r})~,
\end{eqnarray}
where $\rho_0$ is the background value of $\rho$, $R=are^{-\zeta}$, while the density contrast $\delta\equiv\delta\rho/\rho_0$ is calculated as \cite{Harada:2015yda}:
\begin{equation}
\label{density}
    \delta\left({\bf{x}}\right)=\frac{2(1+w)}{5+3w} {\left({\frac{1}{aH}}\right)^2}{\rm{e}}^{2\zeta}\left(\zeta_{,r,r}-\frac{1}{2}\zeta_{,r}^2+\frac{2}{r}\zeta_{,r} \right)~.
\end{equation}
We assume that PBHs are generated in the radiation-dominated era, where $w=1/3$. Substituting Eq. \eqref{density} into Eq. \eqref{deltaM}, it is straightforward to get: 
\begin{eqnarray}
    \label{comp2}
{{\cal C}}(r)&=&\frac{4}{3}r\zeta_{,r}\left(1-\frac{1}{2}r\zeta_{,r} \right)={\cal C}_l(r)\left(1-\frac{3}{8}{\cal C}_l(r) \right)~,
\end{eqnarray}
where we define the linear part of ${\cal C}(r)$: 
\begin{equation}
    \label{comp3}
{\cal C}_l(r)=\frac{4}{3}r\zeta_{,r}~.
\end{equation}

For general non-Gaussian curvature perturbation $\zeta=\zeta(\zeta_G,r)$, the compaction function ${\cal C}(r)$ can be written as a 2D function of $\zeta_G$ and $\zeta_{G,r}$, or equivalently $\zeta_G$ and ${\cal C}_G\equiv r\zeta_{G,r}$ \cite{Gow:2022jfb}. Then ${\cal C}_l(r)$ can be represented by ${\cal C}_G(r)$ as: 
\begin{equation}
    \label{comp4}
    {\cal C}_l(r)=\frac{4}{3}r\zeta_{,r}=\frac{4}{3}r\frac{{{d}}\zeta}{{{d}}\zeta_G}\zeta_{G{,r}}=\frac{4}{3}\frac{d\zeta}{d\zeta_G}{\cal C}_G~.
\end{equation}
Generally speaking, the curvature perturbation (including non-Gaussianity as well) takes the local quadratic form: 
\begin{equation}
    \label{nonG}
    \zeta(\zeta_G)=\zeta_G+(3/5)f_{\rm nl}\zeta_G^2~.
\end{equation}

\begin{figure}
    \centering
    \includegraphics[width=0.6\linewidth]{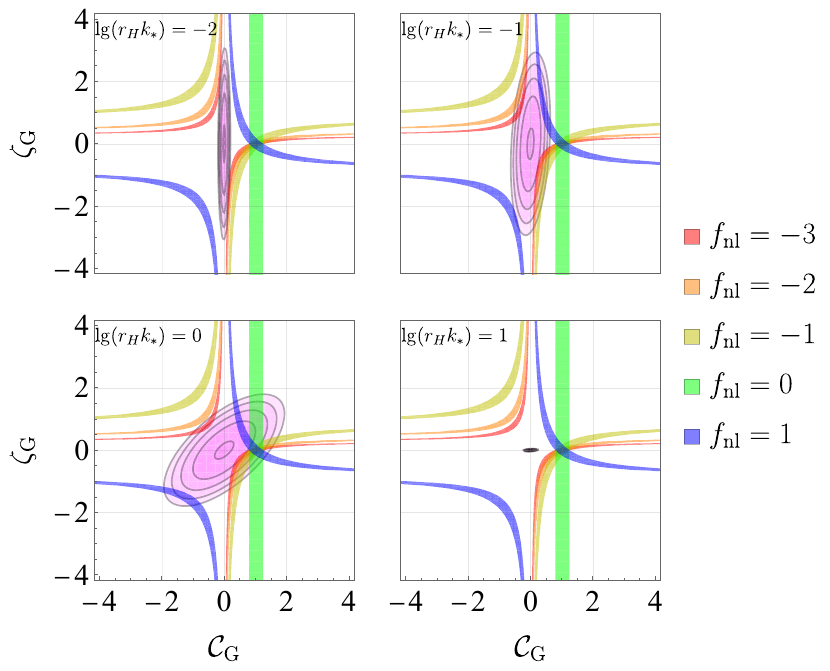}
    \caption{This figure shows the contour lines of PDF of $(\zeta_G, {\cal C}_G)$ with its integral regions under different $r_H$ and $f_\mathrm{nl}$. The elliptical contours, from inside to outside, represent the {\it logarithm} of the probability density, $\lg P(\zeta_G, {\cal C}_G)$, decreasing from $0$ down to $-40$, while color bands represent the integral regions under the corresponding value $f_\mathrm{nl}$.} 
    \label{fig:IntRegionoffPBH}
\end{figure}

The 2D PDF of these two Gaussian variables is 
\begin{equation}
    \label{PDF}
P(\zeta_G, {\cal C}_G) = \frac{1}{2\pi\sqrt{|\bf\Sigma|}}{\rm{Exp}}\left[{-\frac{1}{2}{{\bf{X}}^{\rm{T}}}{{\bf{\Sigma }}^{ - 1}}{\bf{X}}} \right]~,
\end{equation}
with variables matrix and covariance matrix 
\begin{equation}
    \label{RV}
{\bf{X}}\equiv\left[ \begin{array}{l}
\zeta_G\\
{\cal C}_G
\end{array} \right],~~~{\bf{\Sigma}}\equiv\left[ {\begin{array}{*{20}{c}}
{\sigma _{\zeta}^2}&{\sigma _{\zeta{\cal C} }^2}\\
{\sigma _{\zeta{\cal C} }^2}&{\sigma_{\cal C} ^2}
\end{array}} \right]~,
\end{equation}
and the components of the covariance matrix 
\begin{eqnarray}
    \label{Covzetazeta}
\sigma _\zeta ^2 (r_H)&\equiv&\langle\zeta_G^2({\bf x},r_H)\rangle= \int_{}^{}  W^2\left( {k,r_H} \right){\cal T}^2(k,r_H)P_\zeta(k){ d}\ln k~,\\
    \label{CovCzeta}
\sigma _{\zeta{\cal C}}^2(r_H) &\equiv&\langle\zeta_G({\bf x},r_H){\cal C}({\bf x},r_H)\rangle= \frac{4}{3}r_H \int_{}^{}  W\left( {k,r_H} \right)W_{,r_H}\left( {k,r_H} \right){\cal T}^2(k,r_H)P_\zeta(k){ d}\ln k~,\\
    \label{CovCC}
\sigma _{{\cal C}}^2(r_H)&\equiv&\langle{\cal C}^2({\bf x},r_H)\rangle =\frac{16}{9}r_H^2\int_{}^{} W_{,r_H}^2\left( {k,r_H} \right){\cal T}^2(k,r_H)P_\zeta(k){ d}\ln k~.
\end{eqnarray}
Here $W(k,r_H)=\sin(kr_H)/(kr_H)$ is the window function, and the transfer function ${\cal T}(k,r_H)$ is from Eq. \eqref{transfunc}, where we set $\tau=r_H$ for calculating the variance at the horizon crossing of $r_H$. See also \cite{Ando:2018qdb,Young:2019osy} for discussions on various choices of window functions. Generally speaking, for $r_H\rightarrow+\infty$, $\sigma _\zeta,\sigma _{\zeta{\cal C}},\sigma _{{\cal C}}\rightarrow0$. For $r_H\rightarrow0$, $\sigma _{\zeta{\cal C}},\sigma _{{\cal C}}\rightarrow0$ and $\sigma _\zeta={\cal O}(A)$. For $r_H={\cal O}(k^{-1})$,  $\sigma _\zeta,\sigma _{\zeta{\cal C}},\sigma _{{\cal C}}\rightarrow{\cal O}(A)$. Contours of 2D PDFs of $(\zeta_G, {\cal C}_G)$ under different $r_H$ are shown in Fig. \ref{fig:IntRegionoffPBH}. 

A PBH will be formed once the compaction function ${\cal C}$ is over the threshold ${\cal C}_{\rm th}$. Moreover, we restrict ourselves into the Type I perturbation where ${\cal C}_l<4/3$. Then the mass fraction of the PBH is given by: 
\begin{align}
    \label{MassFrac}
\beta(r_H)  = \int_{\cal D} {\cal K}({\cal C} - {\cal C}_{\rm th})^{\gamma_M} P(\zeta_G, {\cal C}_G){d}\zeta_G{ d}{\cal C}_G~,
\end{align}
with integral interval $ {\cal D} = \left\{ {\cal C} > {\cal C}_{\rm th}, {\cal C}_l < 4/3  \right\}$. Eq. \eqref{MassFrac} follows from the empirical formula of the PBH mass \cite{Choptuik:1992jv, Evans:1994pj, Niemeyer:1997mt}:
\begin{equation}
    \label{MassRelation}
M_{\rm PBH}={\cal K}({\cal C}-{\cal C}_{\rm th})^{\gamma_M} M_H~,
\end{equation}
where parameters ${\cal K}$ and ${\gamma_M}$ depend on the equation of state $w$. For the radiation-dominated era where $w=1/3$, we set ${\cal K}=4$ and ${\gamma_M}=0.36$. 
The total abundance of PBHs is then given by the integral over horizon mass $M_H$ (see \cite{Kitajima:2021fpq, Ferrante:2022mui}):
\begin{equation}
        \label{Abund}
        {f_{{\rm{PBH}}}} \simeq\frac{1}{\Omega_{\rm{DM}}}\int_{{M_{{H}}} = 0}^{ + \infty } {{{d \ln}}{M_{{H}}}{{\left( {\frac{{{M_{{H}}}}}{{{2.8\times10^{17}M_\odot }}}} \right)}^{ - 1/2}}\beta}~,
\end{equation}
while the horizon scale $r_H$ and its mass $M_H$ take the relation: 
\begin{equation}
    \label{RadiusandMass}
    {M_{{H}}} \simeq 17{M_ \odot }{\left( {\frac{{{g_ * }}}{{10.75}}} \right)^{ - 1/6}}{\left( \frac{r_H}{10^{-6}{\rm Mpc}} \right)^2}~.
\end{equation}

Although it hardly affects the SIGWs, the non-Gaussianity of the primordial power spectrum will considerably affect the $\beta$ as well as $f_{\mathrm{ PBH}}$ via the compaction function ${\cal C}$ and the domain ${\cal D}$. From Eqs. \eqref{MassFrac} and \eqref{Abund} one can see that, $\beta$ and $f_{\mathrm{PBH}}$ will contain an approximately power-law-like term of $f_{\mathrm {nl}}$, mainly hidden in the compaction function ${\cal C}$. In Fig. \ref{fig:IntRegionoffPBH} we plot the integration region and the PDF as a part of the integrand together, mainly to show the influence of $f_\mathrm {nl}$ to $f_{\mathrm{PBH}}$. The elliptical contours are the hypsographic map of the {\it logarithm} of the probability density, whereas the regions filled with different colors represent the integration domain ${\cal D}$ of Eq.~\eqref{MassFrac}, i.e.\ the overthreshold region, for different values of $f_{\mathrm{nl}}$. Where a region overlaps with the PDF, the PBHs are generated. One can see that both too high and too low $f_\mathrm {nl}$ can push the integration region inward, reach a higher probability distribution and make more PBHs formation. We find $f_\mathrm {nl}\simeq -1$ as a representative value to avoid PBH formation as possible. 

\begin{figure}
    \centering
    \includegraphics[width=0.8\linewidth]{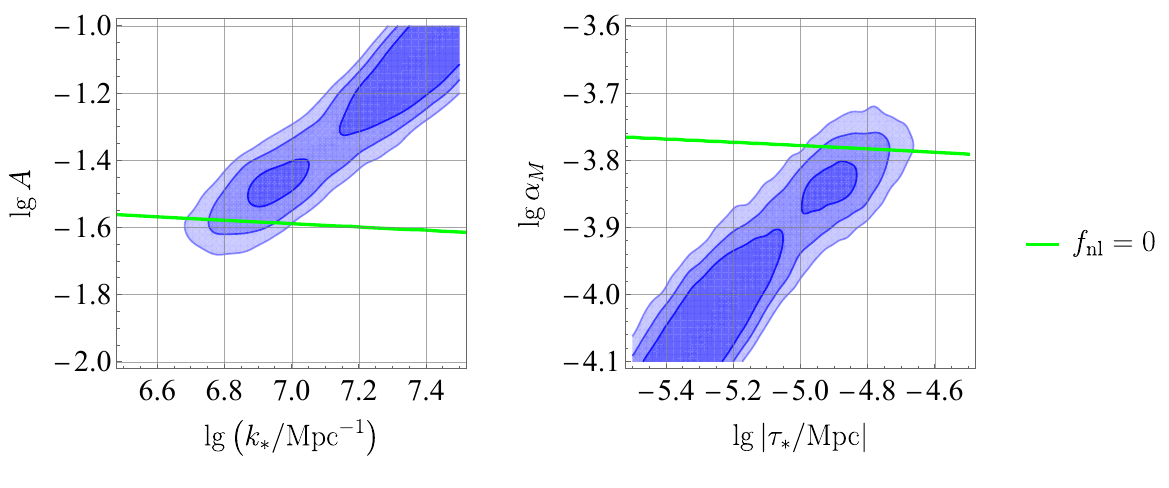}
    \caption{Posterior of BPL parameters (left) and MDR model parameters (right) v.s. $f_{\mathrm{PBH}}=1$ line for $f_{\rm nl}=0$. Region above line in left panel and below line in right panel represents $f_{\mathrm{PBH}}>1$. The contours of the 2D posterior plot from dark to light correspond to the 1$\sigma$, 2$\sigma$, and 3$\sigma$ confidence levels, respectively.} 
    \label{fig:fPBHGauss}
\end{figure}

\begin{figure}
    \centering
    \includegraphics[width=0.75\linewidth]{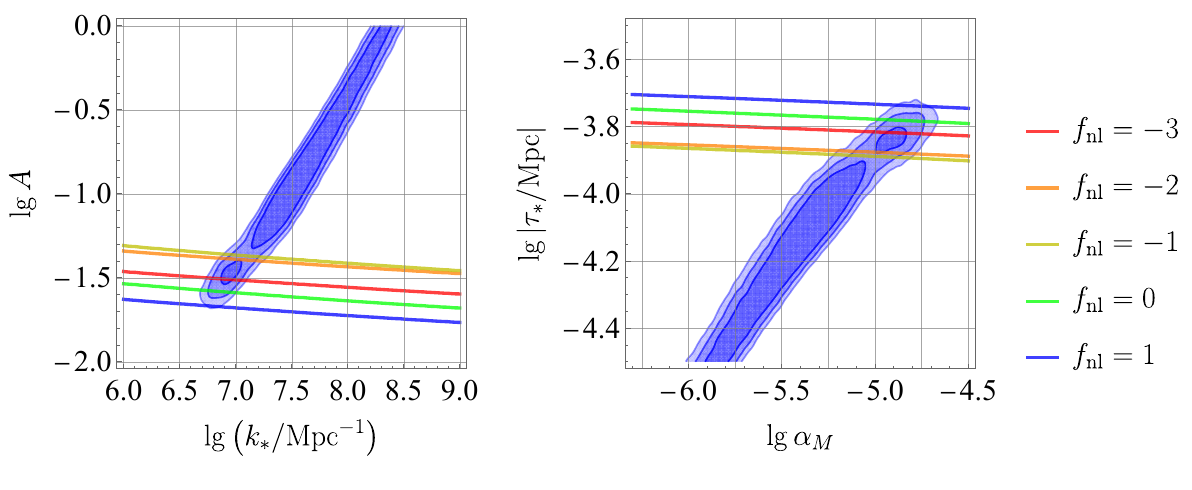}
   \caption{Posterior of BPL parameters (left) and MDR model parameters (right) v.s. $f_{\mathrm{PBH}}=1$ line for more cases with different $f_{\rm nl}$'s. Region above line in left panel and below line in right panel represents $f_{\mathrm{PBH}}>1$.
   } 
    \label{fig:fPBH}
\end{figure}

In Figs. \ref{fig:fPBHGauss} and  \ref{fig:fPBH}  we combine the posterior of parameters from Fig. \ref{fig:SIGW} and Fig. \ref{fig:posteriorMDR} with the $f_{\rm PBH}=1$ line. First of all, to see better the effects of MDR, in Fig. \ref{fig:fPBHGauss} we temporarily turn off the non-Gaussianity by setting $f_{\mathrm{nl}}=0$.  In this figure, the $f_{\rm PBH}=1$ line is {\it marginally} included into the $2\sigma$ region allowed by the PTA data, which means that our model can make contribution to avoiding overproduction of PBH under trivial conditions. Furthermore, as was found in the literature, adding negative non-Gaussianity will give rise to better consequences.
In Fig. \ref{fig:fPBH} we combine the posterior of curvature parameters from Fig. \ref{fig:SIGW} and Fig. \ref{fig:posteriorMDR} with the $f_{\rm PBH}=1$ line for various values of $f_\mathrm{nl}$ together. This figure shows that for positive $f_{\rm nl}$, the $f_{\rm PBH}=1$ line (blue) is out of $2\sigma$ level region, indicating an overproduction problem. Moreover, for the non-Gaussian case with negative $f_{\rm nl}$, the $f_{\rm PBH}$ lines (gold, orange and red) are well within the $2\sigma$ contour, even reaching the edge of $1\sigma$ region. These results hint that both modified dispersion relation and negative non-Gaussianity are helpful in resolving the overproduction problem. This is in agreement with previous results presented in the literature such as \cite{Franciolini:2023pbf, Wang:2023ost, Liu:2023ymk, DeLuca:2023tun, Firouzjahi:2023xke, Chang:2023aba, Pi:2024lsu, Inui:2024fgk, Choudhury:2023fwk, Choudhury:2023fjs}. 

\begin{figure}
    \centering
    \includegraphics[width=0.75\linewidth]{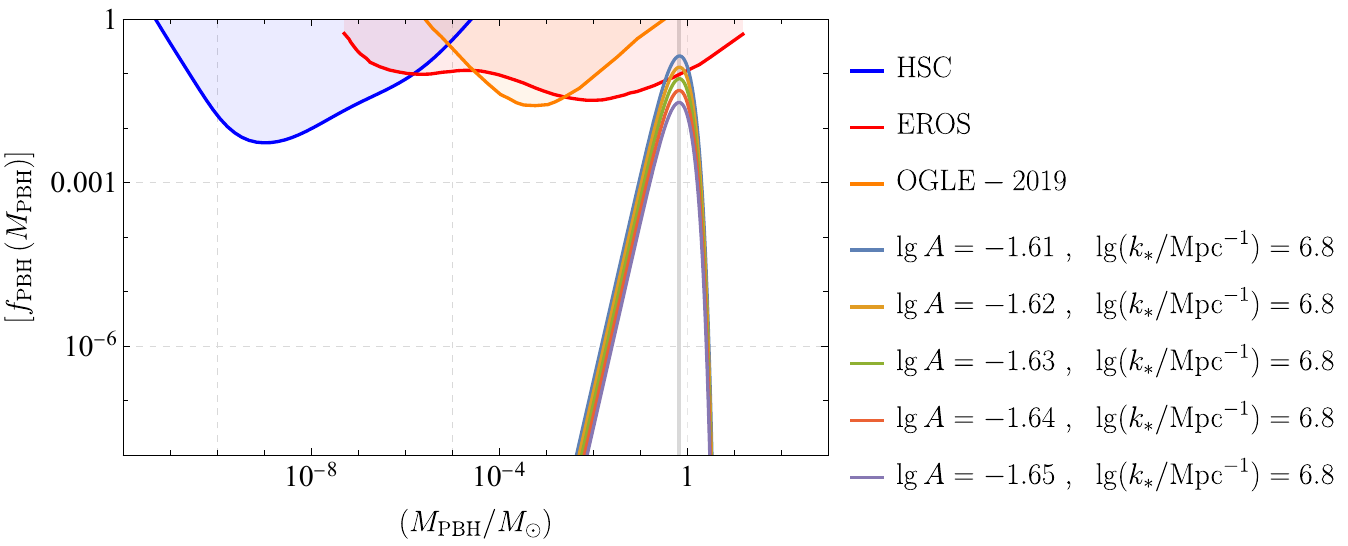}
    \caption{Mass fractions $f_{\mathrm{PBH}}(M_{\rm PBH})$ in logarithmic form, where $M_{\rm PBH}$ is normalized by $M_\odot$. Multiple sets of parameters are used to show that a slight decrease of $\lg A$ can obviously suppress the formation of PBHs, and one can see that $\{-1.63,6.8\}$ is a fine set for $\{\lg A,\lg k/{\rm Mpc^{-1}}\}$ not violating the observation bounds as large.  The shaded regions show the constraints from microlensing observations by the EROS-2 \cite{EROS-2:2006ryy}, the OGLE-2019 \cite{Niikura:2019kqi} and the HSC \cite{Niikura:2017zjd} surveys. }
    \label{fig:fPBHofMPBH}
\end{figure}

One can also obtain the mass distribution of the PBHs, namely $f_{\rm PBH}(M_{\rm PBH})$, with $\int f_{\rm PBH}(M_{\rm PBH})d\ln M_{\rm PBH}=f_{\rm PBH}$. To do this, one needs to perform variable substitution in the integrand from $\left( {{M_{{H}}},{{{\cal C}}_{{G}}},{\zeta _{{G}}}} \right)$ to $\left( {{M_{{\rm{PBH}}}},{M_{{H}}},{\zeta _{{G}}}} \right)$. Combining Eqs. \eqref{MassRelation}, \eqref{comp2}, \eqref{comp3} and \eqref{nonG}, The relation between ${\cal{C}}_{G}$ and $M_\mathrm{PBH}$ can be represented as: 
\begin{align}
    \label{CGofMPBH}
    {{{\cal C}}_{{G}}} = \frac{4}{3} {\left( {\frac{{{{d}}\zeta }}{{{{d}}{\zeta _{{G}}}}}} \right)^{ - 1}}\left[ {1 - \sqrt {1 - \frac{3}{2 }\left[ {{{{\cal C}}_{{\rm{th}}}} + {{\left( {\frac{{{M_{{\rm{PBH}}}}}}{{{{\cal K}}{M_{H}}}}} \right)}^{1/{\gamma_M} }}} \right]} } \right]~.
\end{align}
Thus the mass distribution of PBHs is calculated as:
\begin{eqnarray}
    \label{fPBHofMPBH}
{f_{{\mathrm{PBH}}}}\left( {{M_\mathrm{PBH}}} \right)&=&\frac{1}{{{\Omega _{{\mathrm{DM}}}}}}\int_{{M_{{H}}^{\mathrm{min}}}\left( {{M_{{\mathrm{PBH}}}}} \right)}^{ + \infty }\bigg[{d}\ln {M_{{H}}}{{\left( {\frac{{{2.8\times10^{17}M_\odot}}}{{{M_{{H}}}}}} \right)}^{1/2}}\nonumber\\
&\times& \frac{M_\mathrm{PBH}^2}{{{M_{{H}}}}} \int_\mathbb{R}^{} {\frac{d{\cal C}_G}{dM_\mathrm{PBH}}{\left(\frac{d\zeta}{d\zeta_G}\right)}^{-1}{d}{\zeta _{{G}}}P\left( {{{{\cal C}}_{{G}}},{\zeta _{{G}}}} \right)}\bigg]~,
\end{eqnarray}
where the lower bound of $M_H$ satisfies ${M_{{H}}^{\mathrm{min}}}\left( {{M_{{\mathrm{PBH}}}}} \right)=M_\mathrm{PBH}/({\cal K}{\cal C}_\mathrm{th}^{\gamma_M})$ . In Fig. \ref{fig:fPBHofMPBH} we draw the logarithmic plot of $f_{\rm PBH}$ with respect to $M_{\rm PBH}$ (normalized by $M_\odot$). One can see that under $\lg (k_*/\mathrm{Mpc^{-1}})=6.8$ and $\lg A=-1.63$, $\lg f_{\mathrm{PBH}}(M_\mathrm{PBH})$ reaches its maximum value $-1.09$ for $M_\mathrm{PBH}\simeq 0.67M_\odot$, which means that the PBHs generated in this model will basically have sub-solar mass. Such a fraction can also fit the constraints from observations of microlensing \cite{EROS-2:2006ryy, Oguri:2017ock, Zumalacarregui:2017qqd, Niikura:2019kqi} (see more recent results in \cite{Mroz:2024wia, Mroz:2024mse}) .

\section{conclusions and discussions}
\label{sec:conclusion}
In this paper, we discuss the effect of modified dispersion relation in inflation models on the SIGW and PBH generation, especially the problem of ``overproduction''. 

We start with the very general Mukhanov-Sasaki perturbation equation for inflation models with modified dispersion relation, without denoting specific models. Moreover, $c_s^2$ is assumed to have a suppression at later time in order to make the $k^4$ correction term play an important role. After analytically solving the equation, we find that the power spectrum has a BPL form at the late time of inflation, where the maximum value and the pivot scale are critical for generating PBHs as well as SIGWs. The numerical results of the power spectrum are given in Fig. \ref{fig:curvpert}. 

Such scalar perturbation can induce secondary gravitational waves. With parametrized BPL power spectrum that mimics the numerical result, we obtained the parameter space constrained by the PTA data in Fig. \ref{fig:SIGW}. The result is not too much different from previous works, which indicate that it actually has weak dependence on the power-law indices of curvature spectrum \cite{Franciolini:2023pbf}.   Moreover, we set a dictionary between parameters in BPL parametrization $\{A, k_*\}$ and those in MDR model $\{\alpha_M, \tau_*\}$, and plot the posterior distributions of MDR model parameters in Fig. \ref{fig:posteriorMDR}. We could see that these two sets of parameters are log-linearly related, so these marginals inherit the shape of the corresponding marginals of the BPL parameters. 

To check out whether there is an overproduction problem, we calculate the abundance of PBHs, using the compaction function approach. One can see in Fig. \ref{fig:IntRegionoffPBH} that the overthreshold region depends on the non-Gaussian estimator $f_{\rm nl}$.  In our case, the region for $f_{\rm nl}$ near $-1$ is kept from being too close to the center of the PDF of $(\zeta_G, {\cal C}_G)$, thus is tend to avoid the PBHs being overproduced. Both too high or too low $f_{\rm nl}$ will push the overthreshold region inward to raise the productivity of the PBHs. Moreover, the area of the PDF also changes with the horizon scale $r_H$. In Figs. \ref{fig:fPBHGauss} and \ref{fig:fPBH} we considered impacts on overproduction avoidance both with MDR alone and with non-Gaussianity added. When combined with the SIGW results, we can see in Fig. \ref{fig:fPBHGauss} that $f_{\rm PBH}=1$ line with $f_{\rm nl}\simeq 0$ (no non-Gaussianity) is marginally included into $2\sigma$ region, which means that MDR can be helpful in resolving the overproduction problem. Moreover, the negative non-Gaussianity, if any, will also play an important role.  From Fig. \ref{fig:fPBH} we can see the  $f_{\rm PBH}=1$ line with $f_{\rm nl}\simeq -1$ is well within the $2\sigma$ level. This is in agreement with previous results.

Moreover, we also calculate the mass distribution of the PBHs. We find that in the current case, the mostly generated PBHs can reach a sub-solar mass of ${\cal O}(0.1 \sim 1)M_\odot$ (shown in Fig. \ref{fig:fPBHofMPBH}), with the maximal value of $f_{\mathrm{PBH}}(M_\mathrm{PBH})$ depending on parameters $A$. For overproduction avoidance $f_{\mathrm{PBH}}(M_\mathrm{PBH})$ could reach up to unity, but it can also be lower in order not to be conflicted with constraints from microlensing. Moreover, these constraints might also be released to contain higher $f_{\rm PBH}$ by PBH clustering, by which the influence of PBHs can get reduced (see e.g. \cite{Hawkins:2015uja, Calcino:2018mwh, Belotsky:2018wph}).  

The overproduction problem is an interesting topic about PBH and SIGW generations. It releases the signal that nowadays observations will impose more and more accurate and stringent constraints on these processes. Such constraints will bring challenges to our theoretical analysis and model constructions. However, since we're not very clear about these processes and there are large uncertainties in calculation methodologies remaining, progress on these problems will also deepen our understanding and help us find more optimistic descriptions of them. Therefore, it is interesting to continue discussing about other possibilities on resolving the overproduction problem (as well as other problems), into which we will be devoting ourselves in the future works.

\appendix
\section{Specific inflation model with MDR}
\label{appdx}

In the main text we adopt a Mukhanov–Sasaki equation with a $k^{4}$ correction (Eq. (2.1)) to parameterize the MDR of curvature perturbations during inflation. We use this form as a phenomenological approach rather than starting from concrete models, the purpose of which is to make it cover wider range of application instead of confining into specific cases. Nonetheless, it is easy to understand that such a corrected equation can be derived from theoretical models. We are trying to demonstrate this in the appendix below. 

A simplest correction of a higher-derivative term to a scalar field model has been given in, e.g., Ref. \cite{Gorji:2021isn}, in which the action has been given as
\begin{equation}
    S = \int d^{4}x \sqrt{-g}\left[\frac{M_{p}^{2}}{2}R + P(\phi,X) - \frac{\alpha_M}{2M^{2}}(\Box\phi)^{2}\right],
\end{equation}
where $X = g^{\mu\nu}\partial_{\mu}\phi\,\partial_{\nu}\phi$, $M$ is a cutoff scale, and $\alpha_M$ is a dimensionless constant. The higher-derivative term $(\Box\phi)^{2}$  is a common extension in the framework of field theory. Although sometimes it has been claimed to bring the ``ghost'' degree of freedom, it can be eliminated by the mechanism of spontaneous symmetry breaking of the kinetic term in the Lagrangian, which is also known as ghost condensation \cite{Arkani-Hamed:2003juy, Arkani-Hamed:2003pdi}. 
 
Consider a perturbed FLRW background:
\begin{equation}
    ds^{2} = -(1+2\Phi)dt^{2} + 2a\,\partial_{i}B\,dt\,dx^{i} + a^{2}(1-2\Psi)\delta_{ij}\,dx^{i}dx^{j},
\end{equation}
with the unitary gauge: 
\begin{equation}
     \phi = \phi_{0}(t),~~~ \delta\phi = 0,
\end{equation}
so that all scalar degrees of freedom are absorbed into the metric. In this gauge the curvature perturbation is defined as $\zeta = -\Psi$. Substituting the perturbed metric into the action and expanding to second order, the non‑dynamical fields $\Phi$ and $B$ can be eliminated via the constraint equations. The resulting quadratic action for $\zeta$ (see Ref. \cite{Gorji:2021isn} for the detailed derivation) takes the following form in conformal time $\tau$:
\begin{equation}
    S^{(2)} = \frac{1}{2}\int d\tau\,d^{3}k\;\left[ v_{\mathbf{k}}'^{2} - \left( c_{s}^{2}k^{2} + \tilde{\alpha}_M^{2}k^{4}\tau^{2} - \frac{z''}{z} \right) v_{\mathbf{k}}^{2} \right],
\end{equation}
where $v = z\zeta$ is the Mukhanov–Sasaki variable, $z \propto a$ is related to the background evolution, and
\begin{equation}
    c_{s}^{2} = \frac{P_{X}}{P_{X} + 2X P_{XX}}~,\qquad
\tilde{\alpha}_M = \sqrt{\frac{\alpha_M}{\mathcal{A}}}\,(1-c_{s}^{2})\frac{\dot{\phi}_{0}}{H}~,\qquad
\mathcal{A} = \frac{2\epsilon}{c_{s}^{2}}~,
\end{equation}
with $\epsilon = -\dot{H}/H^{2}$ being the slow‑roll parameter.

In the slow‑roll approximation $z''/z \approx (\theta^{2}-1/4)/\tau^{2}$, and we recover the Mukhanov–Sasaki equation with a $k^{4}$ correction used in the main text:
\begin{equation}
    v_{k}'' + \left( c_{s}^{2}k^{2} + \tilde{\alpha}_M^{2}k^{4}\tau^{2} - \frac{\theta^{2}-1/4}{\tau^{2}} \right) v_{k} = 0~,
\end{equation}
which is nothing but Eq. \eqref{zetabar}. Moreover, one can see that in this model, the coefficient of the $k^4$ term is proportional to $1-c_s^2$, so that in slow-roll I phase this term switches off as $c_s\to1$, while in slow-roll II phase the $c_s^2k^2$ term switches off as $c_s\to0$. This exactly gives rise to Eqs. \eqref{sreq} and \eqref{steq}.

Moreover, such a correction can also arise in the modified gravity model such as $f(T)$ \cite{Qiu:2018nle}, nonlinear models of nonminimal coupling such as Dinkic model \cite{Qiu:2015aha, Qiu:2022klm}. This equation can also be embedded in the effective field theory (EFT) approach, which has been demonstrated in Ref. \cite{Cai:2016thi, Cai:2017tku}. For the EFT action:
\begin{eqnarray}
    S&=&\int d^4x\sqrt{-g}\Bigg[\frac{M_p^2}{2}f(t)R-\Lambda(t)-c(t)g^{00}\nonumber\\
   && +\frac{M_2^4(t)}{2}(\delta g^{00})^2-\frac{m_3^3(t)}{2}\delta K\delta g^{00}-m_4^2(t)(\delta K^2-\delta K_{\mu\nu}\delta K^{\mu\nu})+\frac{\tilde{m}_4^2}{2}R^{(3)}\delta g^{00}\nonumber\\
  &&  -\bar{m}_4^2(t)\delta K^2+\frac{\bar{m}_5(t)}{2}R^{(3)}\delta K+\cdots \Bigg]~,
\end{eqnarray}
The $k^4$ higher-order correction term in the dispersion relation can be sourced by the $\bar{m}_4(t)$ or $\bar{m}_5(t)$ term, namely the $\delta K^2$ and $R^{(3)}\delta K$ operators. This is also proven in Ref. \cite{Ballesteros:2021fsp} with a simpler action. Therefore, Eq. \eqref{zetabar} can be a representative of wide range of cosmological models with MDR studied in the literature.

\begin{acknowledgments}
We thank Shi Pi, Qing-Guo Huang, Xin-zhe Zhang and Hao-Hao Li for useful discussions. This work is supported by the National Science Foundation of China (Grant No. 12575053) and the National Key Research and Development Program of China (Grant No. 2021YFC2203100). 
\end{acknowledgments}









\bibliographystyle{apsrev4-1}
\bibliography{biblio}

\end{document}